# Embarras de richesses in non-DLVO colloid interactions

1. **Forces between solid surfaces in aqueous electrolyte solutions**
   Authors: A.M. Smith, M. Borkovec, and G. Trefalt
   Adv. Colloid. Int. Sci. **275**, 102078, (2020)

2. **Forces between Silica Particles in Isopropanol Solutions of 1:1 Electrolytes**
   Authors: B. Stojimirović, M. Gali, and G. Trefalt
   Phys. Rev. Res. **2**, 023315 (2020)

*Recommended with a Commentary by R. Podgornik*
*School of Physical Sciences, University of Chinese Academy of Sciences, Beijing 100190, China*
*and*
*D. Andelman, School of Physics & Astronomy, Tel Aviv University, 69978 Ramat Aviv, Tel Aviv, Israel*

In its original formulation, the seminal Deryaguin-Landau-Verwey-Overbeek (DLVO) theory of colloidal stability seemed like a simple but realistic description of the world of colloid interactions in electrolyte solutions. It is based on a straightforward superposition of the mean-field Poisson-Boltzmann (PB) electrostatics with the electrodynamic van der Waals (vdW) interactions driven by thermal and quantum fluctuations. However, subsequent developments continued to reveal a much richer and deeper structure of fundamental interactions on the nano- and micro-scale: the granularity and structure of the solvent, charging equilibria of dissociable charge groups, inhomogeneous charge distributions, the finite size of the ions, non-mean-field electrostatics, ion-ion correlations, and more. Today, the original simplicity is gone and we are left with an embarrassingly rich variety of interactions that defy simple classification and reduction to a few fundamental mechanisms. The work of Borkovec and collaborators [1] presents a revealing picture of a more contemporary state-of-the-art picture of colloidal interactions.

In their recent colloidal probe experiment, Stojimirović *et al.* [2] measured forces between silica colloids in isopropanol solutions of 1:1 electrolytes and described how they relate to the aqueous solution experiments [1]. The choice of this particular alcohol solvent is guided by a smaller dielectric constant of isopropanol as compared to water (~18 *vs.* ~80), making the electrostatic coupling in alcohol much stronger than is attainable in aqueous solutions [1]. This high electrostatic coupling unveils a rich phenomenology of colloidal interactions at different salt conditions that are often out of reach in aqueous solutions. In addition to the standard PB double-layer repulsions and vdW fluctuation attractions, such interactions include (i) modification of the Debye screening length characteristic of the double-layer repulsion, (ii) modification of the strength of the double-layer repulsion, (iii) additional monotonic attractive interactions, and (iv) additional non-monotonic interactions. The authors invoked possible mechanisms for each of these effects: ionic pair formation, short-range isopropanol-mediated solvation interaction, charge heterogeneities due to charging equilibria at the surfaces, and the Kirkwood crossover [7] of Coulomb fluids, essentially the transition from the monotonically decaying correlations to a periodic one with a decaying envelope. The described experiments deserve further attention as they highlight several recent developments in measurements of colloidal interactions. Furthermore, they spurred various theoretical developments and conceptual revisions of what we understand to be colloidal interactions.

The force between two spherical silica particles in isopropanol electrolyte solutions was first compared with the predictions of the extended DLVO theory within the local Deryaguin approximation, surmised to be amenable to the superposition of the modified PB and vdW contributions. The general underpinning and limitations of the mean-field PB theory are well known, but more recently, Borkovec and coworkers [1] have set forward the idea that numerical solutions of the full PB equation including charge regulation provide a strikingly accurate quantitative description of the observed colloidal interactions down to a few nm, even in the presence of multivalent ions. This is surprising as for the past two decades the idea that strong coupling approximation should supersede the mean-field approach for multivalent ions has been gaining a foothold in the colloid community. Nevertheless, such an apparent disagreement between the Borkovec's approach and the strong coupling approximation is only superficial.

In fact, the strong coupling suggested by Borkovec and collaborators is relegated to the *charge regulation* of the interacting surfaces [3]. Namely, the adsorption-desorption chemical equilibria of the dissociable surface moieties that can involve monovalent as well as multivalent ions and modify the magnitude of the effective surface charge and surface potential. The strong charge regulation at the interacting surfaces in the presence of multivalent ions can substantially modify the surface charge/potential and, in some cases, even reverse its sign. Furthermore, it may affect the magnitude of the PB interaction and provide the main reason why the PB theory works better for multivalent counterions than for monovalent ions that are not involved in charge regulation.

An additional point to ponder is the range of the PB interactions, quantified by the decay or screening length. There have been previous indications in the context of concentrated aqueous electrolytes that the decay length cannot be always identified as the Debye length. Contrary to the Debye length, the actual decay length increases with a concentration beyond the linearized PB regime. This is presumably due to incomplete salt dissociation with a fraction of salt ions forming Bjerrum pairs, not only in aqueous solutions at an elevated salt concentration (up to 4M) but also in isopropanol salt solutions at much lower concentrations (above 0.1mM). Notably, the Bjerrum pair formation can be also understood as a kind of ionic chemical equilibrium that does not involve surface dissociation, but salt dissociation in the bulk, which connects self-consistently the dielectric decrement of ionic solutions with the screening length [4].

While there is a flurry of theoretical activity in this direction, the Stojimirović *et al.* [2] experiments provide an additional indication, apart from the surface force experiments, on the anomalous decay length variation with salt. Nevertheless, what is still missing to reach a definitive conclusion regarding the effect of Bjerrum pairing theories, is the concurrent measurement of the DC (zero-frequency) dielectric constant of ionic solutions. This can be obtained by taking the zero-frequency limit of the AC dielectric measurements.

A third facet of the extended DLVO theory is the appropriate form of the vdW interaction, usually based on the Lifshitz theory of dipolar fluctuations in dielectric media. Evaluation of vdW interactions requires a full dielectric spectrum of the materials involved in the Matsubara frequency summation that provides the pertaining Hamaker coefficient. In particular, the zero Matsubara frequency is due to the standard thermodynamic fluctuations and can be identified with the Gaussian fluctuations around the mean-field. Several comments are in order here. In aqueous media, the zero frequency term makes the largest contribution to the Hamaker coefficient, being screened with half the Debye length, while in isopropanol solutions, the zero frequency Matsubara term is completely screened already at 5mM salt concentration. More importantly, away from the point of the zero-charge regime of the interacting surfaces, the fluctuations that need to be considered are around the mean electrostatic potential, and not around the zero value of the potential. This is an essential feature of vdW interactions, which nevertheless is rarely taken into account.

Apart from the interactions that can be brought within the extended DLVO paradigm, the non-aqueous isopropanol solutions exhibit also genuine non-DLVO interactions. These interactions are classified as additional repulsions, additional attractions, and non-monotonic oscillatory interactions at higher salt concentration [2], with more or less pronounced similarities to the non-DLVO interactions observed in aqueous media [5]. The non-DLVO short-range repulsion is often associated with the solvent structuring close to the bounding surface. For example, this is the case of hydration repulsion that was extensively modeled and simulated in aqueous solutions, exhibiting a spatial period proportional to the size of the solvent molecules. The oscillatory solvent structural interaction can be nevertheless smoothed out either by the softness of the interacting surface (*e.g.*, phospholipid membrane) or by their roughness (*e.g.*, silica colloidal probe). However, currently, the origin of non-DLVO attraction is not fully understood and is still disputed between different theoretical frameworks.

Apart from the non-mean-field ion-ion correlations and surface charge heterogeneities, an attractive new candidate seems to be the Kirkwood-Shumaker interactions in the case of aqueous media [6]. These interactions are intimately related to the surface adsorption/desorption charge regulation equilibria and can be understood as *monopolar* vdW interactions, to be distinguished from the standard dipolar vdW interactions. The colloidal probe experiments with silica colloids in isopropanol solutions [2] seem to uphold at least to some extent such a monopolar fluctuation mechanism.

Finally, the non-monotonic non-DLVO component seems to be the hardest to disentangle theoretically [2]. Apart from the Kirkwood crossover [7], which was tentatively suggested by Stojimirović *et al*., spatial oscillations in interactions can be due to other mechanisms, of which three lead to qualitatively similar predictions. The first one is based on the non-mean-field ion-ion short-range correlations and steric constraints [8], which implies (on the linearized level) a Poisson equation that is of fourth-order in the derivatives of the potential and leads to a screened oscillatory solution. Apart from the Debye length, this approach introduces another length-scale proportional to the Bjerrum length, setting the crossover between the mean-field and the strong ion-ion correlations regime. Another possibility, formally yielding the same fourth-order Poisson equation, is related to non-electrostatic, *i.e.,* structural spatial variation in the ionic concentration [9]. It leads in general to a more complex behavior than the phenomenological fourth-order Poisson equation with ion-ion short-range correlations, but with qualitatively similar predictions.

As it brings into the fold also the structural non-electrostatic contribution, it could open up new venues to understand the oscillatory colloidal interactions, not necessarily relegated solely to electrostatic effects. We also note that along a similar direction is the *quadrupolar* electrostatics [10], again leading to the fourth-order Poisson or PB equation, with similar consequences for the colloidal interactions. At present, further experimental findings are needed in order to decide which of the theoretical proposals fit most closely the actual physical phenomena as are obtained experimentally.